\documentclass[aps,reprint,twocolumn,a4paper,superscriptaddress]{revtex4-2}

\usepackage{graphicx}
\usepackage{dcolumn}
\usepackage{bm}
\usepackage{setspace}
\usepackage{amsmath}

\begin{document}

\singlespacing

\title{Anharmonic coherent dynamics of the soft phonon mode of a PbTe crystal}

\author{A. A. Melnikov}
\email{melnikov@isan.troitsk.ru}
\affiliation {Institute for Spectroscopy RAS, Fizicheskaya 5, Troitsk, Moscow, 108840 Russia}
\author{Yu. G. Selivanov}
\affiliation {P. N. Lebedev Physical Institute RAS, Moscow, 119991 Russia}
\author{S. V. Chekalin}
\affiliation {Institute for Spectroscopy RAS, Fizicheskaya 5, Troitsk, Moscow, 108840 Russia}

\begin{abstract}

We investigate the ultrafast optical response of PbTe to an intense single-cycle terahertz pulse, resonant with the soft transverse optical (TO) phonon mode of the crystal. We detect multifrequency oscillations of the reflectance anisotropy, which we associate with nonlinear motion of the TO phonon oscillator excited directly by the terahertz pulse. Our observation of monotonically decaying optical anisotropy together with second harmonic oscillations of the TO mode is an evidence of a transient non-centrosymmetric state of the crystal lattice that can be accompanied by the ferroelectric order. We suppose that this state is induced in the PbTe crystal by the intense terahertz pulse via alignment of the local polar nanodomains. This hypothesis is partially supported by the observation of coherent phonons near the satellite phonon mode frequency that are impulsively generated by a femtosecond laser pulse and are considerably enhanced by a synchronous terahertz pulse. 

\end{abstract}

\maketitle

\section{Introduction}

Lead telluride crystals have the NaCl structure and possess a strongly anharmonic transverse optical phonon mode with a rather low frequency near 1 THz \cite{Jantsch}. This mode provides an efficient scattering channel for acoustic phonons and is a prerequisite for low thermal conductivity and good thermoelectric properties of PbTe \cite{Delaire}. Its frequency approximately follows the power law $\omega\sim(T-T_c)^a$, where $a\sim0.5$ and $T_c$ is a critical temperature \cite{Alperin}. Such behavior is typical of the so-called soft phonon modes, which become “frozen” at $T = T_c$ driving a displacive phase transition \cite{Scott}. In the case of lead telluride this is a virtual transition to a lower-symmetry rhombohedral structure, which should exhibit ferroelectric order, but is never realized for undoped (binary) crystals. The critical temperature $T_c$ is negative and PbTe remains paraelectric even at temperatures close to zero. Therefore, lead telluride is referred to as incipient ferroelectric \cite{Jiang1}.

Bringing a PbTe crystal closer to this transition is of interest from the point of view of fundamental physics, as well as practical applications, such as control of ferroelectric polarization and increasing the thermoelectric figure of merit. The energetic proximity to a non-centrosymmetric structure and the softness of the TO phonon mode make lead telluride crystals particularly susceptible to various external stimuli including strain \cite{Murphy}, magnetic field \cite{Baydin}, light-induced injection of charge carriers \cite{Jiang2}. Methods of materials engineering can be also applied. Replacement of several percent of Pb atoms by Ge or Sn makes PbTe crystals ferroelectric at lower temperatures \cite{Jantsch}. Ferroelectricity of intrinsic lead telluride in the 2D limit was theoretically predicted \cite{Zhang}.

Since the TO mode is infrared active, it is possible also to explore a direct approach --- excitation of atomic vibrations by terahertz radiation \cite{Li, Nova}. Here we report a study of the ultrafast optical response of a PbTe crystal to an intense picosecond terahertz pulse resonant with the TO phonon mode. We have found that such a pulse induces oscillations of reflectance anisotropy of the crystal at three distinct frequencies, which are related approximately as 1:2:3 and decrease accordingly upon cooling. We associate these oscillations, respectively, with nonlinear coherent TO vibrations of atoms of PbTe and with the second and third harmonics of this motion. The oscillations are superimposed onto a monotonically decaying transient, which we interpret as a short-lived symmetry lowering of the crystal caused by alignment of the polar local nanodomains by the intense terahertz pulse. Though anisotropy of reflectance not necessarily implies non-centrosymmetry of a crystal, the fact that it is the soft phonon mode that is driven by the terahertz pulse and the observation of its second harmonic make it possible to conjecture that this nonequilibrium state lacks center of inversion and is accompanied by the ferroelectric order. 

In addition, we use near infrared femtosecond laser pulses to generate in the PbTe crystal coherent phonons at a frequency of $\sim$ 1.4 THz ($\sim$ 6 meV), which behave differently with respect to the observed harmonics. This phonon mode can be related to the satellite peak of the main TO phonon of PbTe that has been detected or theoretically predicted is several previous studies \cite{Burkhard, Jensen, Li2, Li3, Chen, Ribeiro, Guzelturk, Li4}. In some of those papers this additional phonon mode was associated by the authors with a local symmetry lowering due to off-centering of the Pb atoms and formation of correlated dipoles \cite{Delaire, Jensen, Guzelturk, Bozin, Sangrigorio}. We have found that an additional intense terahertz pulse synchronous with the femtosecond pump pulse considerably increases the amplitude of these coherent phonons. This fact is in agreement with our hypothesis that the terahertz pulse aligns these local dipoles inducing a macroscopic transient state with a lower symmetry.

\section{Experimental details}

The crystal of lead telluride that was studied in the present work was grown by the Czochralski method and cleaved along the (100) plane. The room-temperature concentration of charge carriers was 3$\cdot$10$^{18}$cm$^{-3}$ (p-type). Measurements of the ultrafast dynamics were performed using the standard pump-probe layout. Pump terahertz pulses were generated via optical rectification of femtosecond laser pulses with tilted front in a crystal of lithium niobate \cite{Stepanov}. For that purpose, we used a fraction of the output beam from a regenerative Ti:sapphire amplifier (Spitfire HP, Spectra Physics) --- 1.8 mJ pulses with a duration of 70 fs, central wavelength of 800 nm and a repetition rate of 1 kHz. After collimation and focusing by parabolic mirrors the resultant terahertz pulses had a peak electric field of about 500 kV/cm, a duration of about 1 ps, while their peak frequency was near 1 THz (see Fig. 1(a) and Fig. 1(b)). Pump pulses at the central wavelength of 1.3 $\mu$m and with a duration of $\sim$ 70 fs were generated by an optical parametric amplifier (TOPAS, Light Conversion).

In order to characterize the excited state created by the pump pulse in the PbTe crystal we detected anisotropy of reflectance using weak probe femtosecond pulses at 800 nm. To do that we measured rotation of polarization of a probe pulse reflected from the sample using a Wollaston prism and a pair of amplified photodiodes. The initial polarization of the probe beam was set to 45$^\circ$ relative to the vertical polarization of the terahertz pulses. The measurement was repeated multiple times for open and closed pump beam (modulated by an optical chopper) and the normalized difference value was calculated and then averaged. The resulting quantity will be denoted as $\varphi$ throughout the text. Due to this normalization procedure, it is close to zero in equilibrium. Pump and probe beams were incident onto the sample at an angle of $\sim15^\circ$. Low temperature measurements were performed with the PbTe crystal placed inside a liquid nitrogen cryostat with optical windows made from TPX polymer.

\section{Results and discussion}

\begin{figure}
\begin{center}
\includegraphics{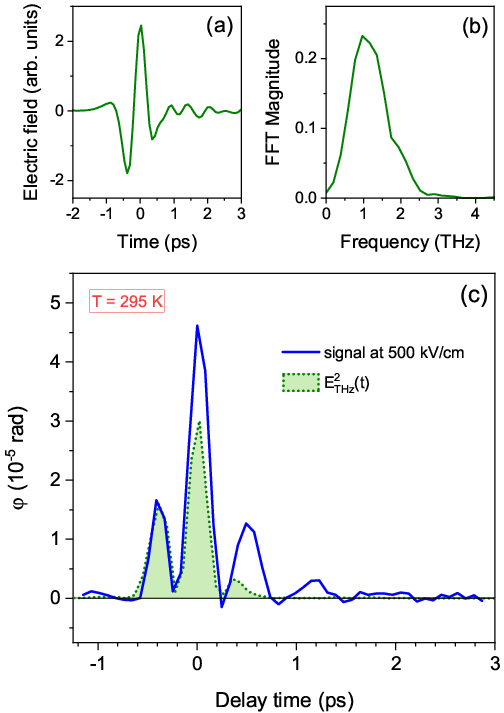}
\end{center}
\caption{\label{fig1} (a) -- Temporal profile of the electric field of the pump terahertz pulse. (b) -- Spectrum of the pump terahertz pulse. (c) -- Polarization rotation $\varphi$ of the probe pulse reflected from the PbTe crystal as a function of the pump-probe delay time measured at room temperature (blue solid line). Normalized squared electric field of the pump terahertz pulse in arbitrary units (green dotted line with filled area).}
\end{figure}

Figure 1(c) shows temporal evolution of the polarization rotation $\varphi$ of the probe pulse during and after the illumination of the PbTe crystal by the pump terahertz pulse. The initial part of the signal is somewhat similar to the squared electric field of the pulse and is likely due to nearly instantaneous Kerr effect of electronic origin. This waveform is followed by strongly damped oscillations superimposed onto a monotonic transient. An interesting property of this compound trace is that its minima caused by the oscillations are always near zero, while the resulting values are all positive (neglecting the noise). This can be realized only if the decay times of the monotonic component and of the oscillations are rather close or identical. This feature is illustrated in Fig. 2, where the signal $\varphi(t)$ is fitted by the function $A_0e^{-t/\tau}(1-\sin(2\pi\nu t-\phi))$ starting from $t$ = 250 fs, when the pump pulse is almost over. Here $\tau$ = 0.5 ps, $\nu$ = 1.6 THz and $\phi$ = 0.8 are the decay time, the frequency and the phase of the oscillations, respectively. 

Oscillations in an optical response of a crystal to an ultrashort laser pulse can often be explained by the impulsive generation of coherent phonons \cite{Dekorsy}. In this case collective vibrations of atoms periodically modulate the refractive index of the crystal (if allowed by symmetry) and can be detected by optical probing. Then the apparent frequencies of oscillations in the decay traces correspond to the frequencies of particular phonon modes. The inset to Fig. 2 shows spectra of three traces from the main panel that were recorded in three consecutive measurements. Two well defined peaks are visible at $\sim$ 1.6 THz and $\sim$ 2.6 THz. These values should be compared to the known frequencies of lattice vibrations of PbTe crystals.

The frequencies of the TO and LO phonon modes of PbTe at room temperature that can be found in the literature usually lie near 1 THz and 3.3 THz, respectively (see e. g. \cite{Jantsch, Alperin, Cochran, Guzelturk}). An additional satellite peak near 1.4--1.6 THz was also reported in a number of papers \cite{Burkhard, Delaire, Jensen, Li2, Li3, Guzelturk, Li4}. However, its origin is still not clear (see the discussion below). It should be noted that frequency values obtained in different studies vary, which is particularly true for the TO mode and is likely due to different concentrations of free charge carriers in the crystals under study. Herein we will refer to one of the classical works on inelastic neutron scattering on single crystals of PbTe for temperature dependent values of the TO phonon frequency \cite{Alperin}. The crystal of PbTe studied in that work was grown by the Czochralski method and was p-type with carrier concentration of (3--5)$\cdot$10$^{18}$cm$^{-3}$, similar to our crystal. The TO frequency measured by the authors at room temperature was 0.87 THz. Thus, we can conclude that the oscillation frequencies detected in our measurements at 295 K do not match the standard phonon frequencies reported for PbTe crystals.

Moreover, symmetry considerations impose restrictions on the types of coherent phonons that can be detected using our experimental layout. Rotation of polarization of the probe pulse upon reflection from the excited sample implies a certain anisotropy of reflectance induced by coherent motion of atoms. Therefore, in order to be detected, coherent phonons should modulate polarizability of the crystal and be Raman active. However, it was shown theoretically that for crystals with rock-salt structure there is no first-order Raman scattering \cite{Born, Ferraro}. The second order effect is generally much weaker, and thus, for a perfect PbTe crystal we should not expect oscillations in the optical anisotropy response to an ultrashort pulse of moderate intensity. 

\begin{figure}
\begin{center}
\includegraphics{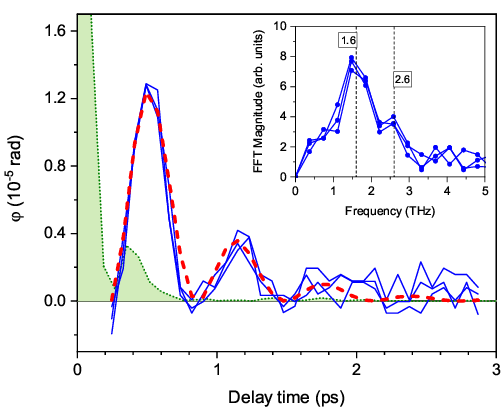}
\end{center}
\caption{\label{fig2} A magnified view of the signal $\varphi(t)$. The main panel shows three traces measured in three consecutive scans (blue solid lines). The average of these traces was shown in Fig. 1(c). The red dashed line represents a fit according to the formula $A_0e^{-t/\tau}(1-\sin(2\pi\nu t-\phi))$ (see text). The inset shows spectra of the three traces from the main panel. The two most prominent peaks are indicated by vertical dashed lines and the corresponding frequency values in THz.}
\end{figure}

In order to clarify the origin of the observed oscillations, we have repeated measurements at lower temperatures. The results obtained at 85 K for two values of the peak electric field of the pump terahertz pulse are shown in Fig. 3. We first note that the above-mentioned property of the signal measured at 295 K, namely the closeness of the decay times of the monotonic component and of the apparent oscillations, is in general retained at low temperatures. However, as can be seen from the spectra of the signals in Fig. 3(b), the two peaks observed at room temperature shift to lower frequencies upon cooling, while a distinct new peak appears. An interesting fact is that the frequencies of these three peaks are related approximately as 1:2:3. The intensity of the first peak at $\omega_1$ = 0.72 THz increases approximately linear with the peak electric field of the pump pulse, while the second one at $\omega_2$ = 1.37 THz demonstrates clearly nonlinear growth. The same can be said about the third peak at $\omega_3$ = 2.19 THz only tentatively due to a lower signal-to-noise ratio in this case.

\begin{figure}
\begin{center}
\includegraphics{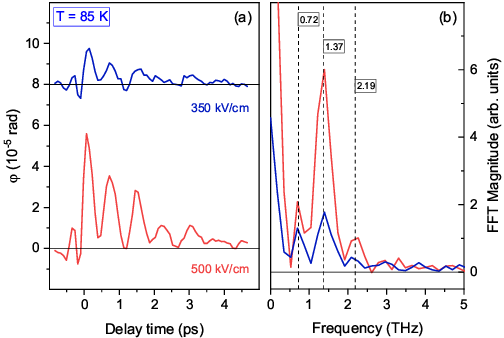}
\end{center}
\caption{\label{fig3} (a) -- Two traces $\varphi(t)$ measured using pump terahertz pulses with 350 kV/cm (upper curve) and 500 kV/cm (lower curve) peak electric field strength. The upper curve was shifted vertically for clarity. (b) -- Spectra of the signals shown in panel (a). Three spectral peaks are indicated by vertical dashed lines with the frequency values shown in THz.}
\end{figure}

Normalized spectra of the oscillations obtained at temperatures of 85, 120, 140, 170, and 200 K are presented in Fig. 4(a). In this case, in order to reduce noise, we performed fast Fourier transform not on the original signal, but on the autocorrelation of its oscillating part. This procedure reproduces the frequencies of the oscillatory components but effectively squares their relative amplitudes. It can be seen that all three spectral peaks, $\omega_1$, $\omega_2$, and $\omega_3$, shift to lower frequencies upon cooling. A certain narrowing of the lines is also observed, most consistently for the most intense line at $\omega_2$. In Fig. 4(b) we compare the recorded temperature dependence of $\omega_1$, $\omega_2$, and $\omega_3$ with the corresponding dependence of frequencies of the TO phonon mode, as well as of its second and third harmonics, calculated from the data presented in \cite{Alperin}. The agreement is rather good if we neglect a certain systematic deviation for  $\omega_3$ at lower temperatures. We have also compared the decay rate of the monotonic component of the reflectance anisotropy signal $1/2\tau_1$ with the damping rate $\gamma$ of the most intense oscillatory component at  $\omega_2$ (assuming the Lorentzian line shape). The value of $\tau_1$ was determined by fitting the signal with a single exponential function, while $\gamma$ was estimated as FWHM of the corresponding spectral line. Again, we can see that these values are close in the whole temperature range of measurements (see Fig. 4(c)).

\begin{figure}
\begin{center}
\includegraphics{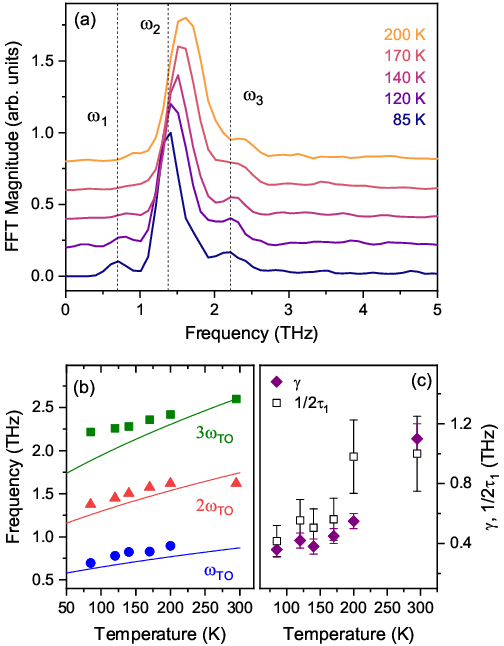}
\end{center}
\caption{\label{fig4} (a) -- Spectra of the autocorrelations of the oscillating components of the terahertz-induced signals $\varphi(t)$, measured at temperatures 200 K, 170 K, 140 K, 120 K, and 85 K (top -- bottom). The curves were normalized to the maximum of the most intense peak and shifted vertically for clarity. Vertical dashed lines indicate positions of the three spectral peaks $\omega_1$, $\omega_2$, and $\omega_3$ in the spectrum obtained at 85 K. (b) -- Temperature dependence of $\omega_1$ (circles), $\omega_2$ (triangles), and $\omega_3$ (squares). Solid lines illustrate temperature dependence of the frequency of the TO phonon mode of PbTe, as well as of its integer multiples $2\omega_{TO}$ and $3\omega_{TO}$. The corresponding curves were calculated using values of $\omega^2_{TO}$ from \cite{Alperin}. (c) -- Temperature dependence of the damping rate $\gamma$ of oscillations at $\omega_2$ (diamonds), evaluated as FWHM of the corresponding peak in the spectra shown in panel (a). Empty squares indicate values of the decay rate $1/2\tau_1$ of the monotonic exponential component of the signal for each temperature.}
\end{figure}

The experimental results described above can be tentatively interpreted as follows. First, we identify the observed spectral peaks at frequencies $\omega_1$, $\omega_2$, and $\omega_3$ with the fundamental frequency of the TO phonon mode of the PbTe crystal, and with its second and third harmonics, respectively. The intense terahertz pulse is resonant with the infrared active TO mode and drives the atomic motion directly, creating a coherent wave packet with a rather large amplitude. This process can in principle be illustrated as motion of a classical oscillator in a potential that is strongly anharmonic due to anharmonicity of the crystal lattice of PbTe and the “softness” of this TO mode. The oscillator is nonlinear and the spectrum of its vibration contains harmonics of the fundamental frequency \cite{vonHoegen}. The second order nonlinearity not only enables generation of the second harmonic but also “rectifies” atomic oscillations thereby inducing a transient distortion of the crystal lattice \cite{Först, Kahana}. The decay of this distortion is monotonic and its characteristic time is equal to the decay time of the oscillations at $\omega_2 = 2\omega_{TO}$. 

This simplified model can account for the basic features of the detected reflectance anisotropy signal. However, it contains several inconsistencies. First, the model implies that the damping rate $\gamma$ of the second harmonic oscillations is twice larger than the damping rate $\gamma_{TO}$ of the vibrations at the fundamental frequency $\omega_{TO}$. Using this relation, we can obtain from our data values of $\gamma_{TO}$ from $\sim$ 0.2 THz to $\sim$ 0.6 THz in the temperature range 100--300 K, which are considerably larger than the typical values available in the literature (see e. g. \cite{Burkhard2, Baydin} for a typical ratio $\gamma_{TO}/\omega_{TO}\sim0.1$). More importantly, the considerable second-order nonlinearity of atomic oscillations implies that the cubic symmetry of the crystal is lowered immediately after the action of the terahertz pulse and a transient non-centrosymmetric structure is created. It is not clear, however, by which mechanism the coherent TO phonons can break the inversion symmetry of the rock-salt structure of PbTe by themselves.

Here we note that in several recent studies on x-ray and neutron scattering of PbTe it was argued that though the “average” structure of the crystal is cubic, on the scale of several unit cells there occurs dynamic off-centering of Pb atoms, which leads to a local symmetry lowering and formation of non-centrosymmetric polar nanoregions \cite{Jensen, Bozin, Sangrigorio}. It is possible to suppose that an intense terahertz pulse aligns these fluctuating local dipoles and a short-lived “macroscopic” excited state is formed lacking center of inversion \cite{Cheng}. In this case the coherent TO phonons generated by the same terahertz pulse and their harmonics will be visible in the reflectance anisotropy signal as long as this dipole-aligned state exists. The decay of the latter then corresponds to the observed monotonic relaxation and the damping of the oscillations at $\omega_2 = 2\omega_{TO}$. 

A notable feature that was associated with the polar nanodomains in the above-mentioned studies was a satellite peak that appeared for temperatures above $\sim$ 100 K at slightly higher frequencies relative to the TO phonon. This additional phonon mode at frequencies $\sim$ 1.4-1.6 THz was also observed experimentally and modeled numerically in several previous works \cite{Burkhard, Delaire, Jensen, Li2, Li3, Guzelturk, Li4, Chen, Ribeiro}. An alternative interpretation of this mode was suggested, according to which it results from anharmonic lattice effects, such as the interaction of optical and acoustic phonon modes \cite{Burkhard, Li2, Li3, Chen}. However, the reported temperature-dependent behavior of the peak varies from study to study and its origin still remains unclear.

\begin{figure}
\begin{center}
\includegraphics{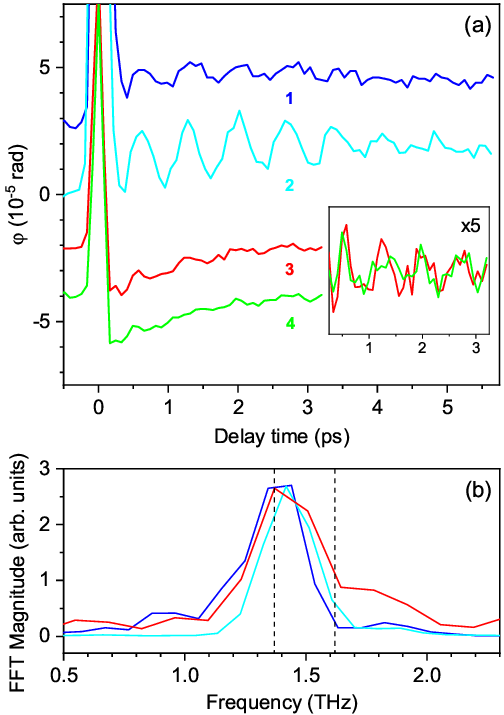}
\end{center}
\caption{\label{fig5} (a) -- Polarization rotation $\varphi$ of the probe pulse reflected from the PbTe crystal as a function of the pump-probe delay time measured under following conditions: 85 K, 1.3 $\mu$m pump (curve 1); 85 K, 1.3 $\mu$m pump with the synchronous terahertz pulse (curve 2); 295 K, 1.3 $\mu$m pump (curve 3); 295 K, 1.3 $\mu$m pump with the synchronous terahertz pulse (curve 4). The energy density of the pump pulse at 1.3 $\mu$m was $\sim$ 0.5 mJ/cm$^2$. The curves were shifted vertically for clarity. The small monotonic background in all four traces is caused by large ($\sim10^{-3}$) isotropic reflectance changes induced by the pump pulse and detected here due to polarization leakage. The inset shows a comparison of the oscillations in the room-temperature traces 3 and 4. The view is magnified by a factor of 5. (b) – Normalized spectra of the autocorrelations of the oscillating components of the traces 1, 2, and 3. Vertical dashed lines indicate the position of the $\omega_2$ peak at 85 K (left line) and at 295 K (right line).}
\end{figure}

We have performed additional measurements of the photoinduced reflectance anisotropy of PbTe with femtosecond pump pulses at 1.3 $\mu$m and detected very weak oscillations at $\Omega\sim$ 1.4 THz, the behavior of which differs considerably from the behavior of the coherent TO phonons and their harmonics generated by the intense terahertz pulse. The results are shown in Fig. 5. The amplitude of these oscillations increases almost twofold upon lowering the temperature from 295 K to 85 K, while their frequency and linewidth decrease only slightly. The observed frequency shift is, however, considerably smaller than that of the second harmonic of the coherent TO phonons (see Fig. 5(b)). We have also found that at 85 K a synchronous excitation of the crystal by a terahertz pulse with the peak electric field of $\sim$ 500 kV/cm leads to a further increase of the amplitude of the oscillations approximately by a factor of three (here synchronization means temporal coincidence of the femtosecond pulse with the peak electric field of the terahertz pulse). In this case a small high-frequency shift of the corresponding spectral line is observed. At the same time, at room temperature the effect of the additional terahertz excitation can be regarded as negligibly small.  

The observed oscillations at $\Omega\sim$ 1.4 THz are likely caused by coherent phonons impulsively excited in the PbTe crystal by the 1.3 $\mu$m femtosecond laser pulse. We suppose that these phonons can be associated with the fluctuating polar nanodomains of PbTe by analogy with the previous studies, in which the satellite peak of the TO phonon mode was detected. First, such interpretation could account for the small amplitude of the oscillations. Indeed, as we have already noted above, since the static (or dynamic “averaged”) crystal structure of PbTe is of the cubic rock-salt type, generation and anisotropic detection of coherent phonons are forbidden to the first order in phonon coordinates. Local symmetry lowering can lift this selection rule and allow generation of weak coherent phonons. Second, as we assumed above, an intense terahertz pulse could align the local polar nanoregions and create a transient state with a lower symmetry and enhance the oscillations at $\Omega\sim$ 1.4 THz. Such effect is indeed observed in our experiments (see Fig. 5). 

The effect of cooling on these features is, however, somewhat controversial, if we refer to the studies, in which the satellite peak was reported, staying at the same time in the framework of our hypothesis. In our case both the amplitude of coherent phonons at $\Omega\sim$ 1.4 THz and the effect of the terahertz pulse on this amplitude increase upon cooling, while the frequency shift is relatively small. However, in most of the neutron and x-ray scattering experiments the amplitude and the frequency of the satellite peak increased considerably upon heating starting from $\sim$ 100 K \cite{Delaire, Jensen, Li2, Li4}. Nevertheless, recently there was a study, in which this peak was observed using inelastic neutron scattering even at temperatures down to 5 K \cite{Li4}. One numerical study \cite{Ribeiro} and a recent paper on ultrafast terahertz emission \cite{Guzelturk} from PbTe reported negligible or small frequency shifts of the satellite peak upon lowering the temperature, similar to our work. It should be noted also that the specific temperature behavior observed in our case may be caused by the inherent properties of the impulsive mechanism of coherent phonon generation by an ultrashort laser pulse, which is essentially different from the x-ray or neutron scattering on elementary excitations. Therefore, the partial disagreement of our results with several structural studies does not necessarily contradict our interpretation.

\section{Conclusion}

In summary, we have studied the optical response of PbTe to an intense nearly single-cycle terahertz pulse. Relaxation of the induced anisotropy of reflectance of the crystal was accompanied by oscillations at three distinct frequencies, which we associated with the fundamental frequency of the TO phonon mode and with its second and third harmonics. All three spectral peaks shifted to lower frequencies upon cooling in accordance with the well-established temperature dependence of the soft TO mode of PbTe. Measured decay traces contained a monotonic exponential transient, the characteristic time of which was close to the lifetime of the second harmonic oscillations in the whole temperature range covered. Excitation of the PbTe crystal by a femtosecond pulse at 1.3 $\mu$m generated coherent phonons at the frequency of $\sim$ 1.4 THz, close to the well-known but controversial satellite of the TO phonon mode. The damping of these oscillations, as well as the low-frequency shift and line narrowing upon cooling were considerably smaller than those of the second harmonic of the TO phonon mode, while at low temperature they could be enhanced by a synchronous terahertz pulse. In order to interpret the obtained results, we supposed that the terahertz pulse aligns fluctuating local polar nanoregions of PbTe creating a macroscopic non-centrosymmetric state of the crystal lattice. In this transient state the nonlinear motion of the TO phonon oscillator, resonantly excited by the same pulse, becomes detectable due to lifting of the selection rules. The alignment of local dipoles also enhances impulsive coherent excitation of the associated satellite phonon mode by a femtosecond laser pulse.

\begin{acknowledgments}

The reported study was funded by the Russian Science Foundation, Project No. 23-22-00387. The authors are grateful to Kirill Boldyrev for help with low temperature measurements. 

\end{acknowledgments}

\end{document}